\documentclass[%
 reprint,
 amsmath,amssymb,
 aps,
]{revtex4-1}
\usepackage{color}
\usepackage{graphicx}
\usepackage{dcolumn}
\usepackage{bm}

\begin{document}

\preprint{APS/123-QED}

\title{The energy cascade at the turbulent/non-turbulent interface}

\author{Y. Zhou$^{1}$} \author{J. C. Vassilicos$^{2}$}%
\email{john-christos.vassilicos@centralelille.fr}
\affiliation{$^{1}$School of Energy and Power Engineering, Nanjing
  University of Science and Technology, Nanjing 210094,
  China\\ $^{2}$Univ. Lille, CNRS, ONERA, Arts et Metiers Institute of
  Technology, Centrale Lille, UMR 9014 - LMFL - Laboratoire de
  M\'ecanique des fluides de Lille - Kamp\'e de Feriet, F-59000 Lille,
  France
}%

\date{\today}

\begin{abstract}
Interscale energy transfers at the vicinity of the
turbulent/non-turbulent interface are from small to large scales in
directions close to the interface's tangent plane where motions are
predominantly stretching, but from large to small scales in the other
directions where motions are predominantly compressive and
significantly correlated with square angular momentum. An important
role in this predominance is played by the extreme compressive motions
which can be significantly more likely than extreme stretching motions
even where motions are on average stretching. The most intense
interscale transfer rates and dissipation occur when the interface is
as far as possible from the wake centreline.

\end{abstract}

\maketitle


Typically for statistical physics, the number of degrees of freedom in
fully resolved Navier-Stokes simulations of turbulent flows is
prohibitively large at high Reynolds numbers (ratio of inertial to
viscous forces). The turbulence problem is to reliably reduce the
number of degrees of freedom, either universally or in different ways
in different universality classes. This reduced order modeling
requires understanding of the non-linear physics which determine
turbulence dynamics and statistics.

An important feature of many environmental, geophysical and industrial
turbulent shear flows (turbulent wakes, jets, boundary/mixing layers,
etc) \textcolor{black}{of very wide relevance, including for mixing and
  cloud physics \cite{SILVA-ARFM, MELLADO-ARFM}}, is the presence of a
sharp interface between potential non-turbulent flow and vortical
turbulent flow (first studied systematically by
\citet{Corrsin}). These Turbulent/Non-Turbulent Interfaces (TNTI) pose
a serious challenge to reduced order modeling because they are
conceptually hard to reconcile with eddy viscosities \cite{Westerweel}
and they incorporate a wide range of scales of motion. Intimately
linked to this wide range of scales which characterise turbulent flows
is the energy cascade which is a direct consequence of non-linearity
at high Reynolds numbers. To the authors's knowledge, nothing is known
about the energy cascade and interscale energy transfers at the
TNTI. Are these interscale transfers from large to small or from small
to large scales, and do they depend on orientation relative to the
TNTI?  How do they correlate with local rotation, local stretching and
compression, and perhaps also other non-linear processes such as
turbulent transport?  Do interscale transfers and turbulence
dissipation depend on TNTI location?  This letter offers answers to
these questions which concern essential turbulence physics of direct
relevance to reduced order modeling, in particular coarse-grained
representations of turbulence and related subgrid scale modeling for
Large Eddy Simulations where interscale energy transfer is key.

We use data from a state-of-the-art massively parallel Direct
Numerical Simulation (DNS) of spatially evolving axisymmetric and
incompressible turbulent wake obtained by \citet{Dairay}. In terms of
free stream velocity $U_{\infty}$ and area $L_{b}^{2}$ of the bluff
plate placed normal to the incoming laminar free stream, the Reynolds
number is $U_{\infty} L_b/\nu = 5000$ (fluid's kinematic viscosity
$\nu$). The computational domain is long \textcolor{black}{($120L_b$)}
and data were reliably collected up to a distance $x=100L_b$ from the
plate. The spatial resolution increases with $x/L_b$ and equals the
Kolmogorov microscale $\eta_K$ at $x= 60L_b$ where our study's results
were obtained. \textcolor{black}{At $x=60L_b$ on the centreline,
  $\eta_K \approx 0.025 L_b$ and the Taylor and longitudinal integral
  length scales are $\lambda \approx 0.37 L_b$ and $L \approx 1.97
  L_b$.}

We calculate statistics conditional on TNTI location within the $x=
60L_b$ plane. We locate the TNTI in the way that \citet{Zhou} located
it for the same DNS data set, by computing the instantaneous area
$A_t$ within the $x= 60L_b$ plane where the modulus $\omega$ of the
instantaneous enstrophy exceeds a threshold $\omega_{th}$. $A_t$
decreases with increasing $\omega_{th}$, reaching zero at $\omega_{th}
= \omega_{max} (t)$, the maximum $\omega$ in the $x= 60L_b$ plane at
time $t$. The presence of the TNTI manifests itself by a plateau in
the plot of $A_t$ versus $\omega_{th}/\omega_{max}$ (this plot is
similar whether $\omega_{th}/\omega_{max}$ is kept constant in time or
not), see figure 1 in \citet{Zhou}.  At $x= 60L_b$, $A_t$ is about
constant over the $\omega_{th}/\omega_{max}$ range from about
$2.5\times 10^{-4}$ to $10^{-2}$. This wide vorticity range is
taversed over a thin region of space, the TNTI, which explains the
wide plateau in the $A_t$ versus $\omega_{th}/\omega_{max}$ plot. In
this letter, we chose $\omega_{th}/\omega_{max} = 4\times 10^{-4}$ to
locate points on the TNTI in the $x= 60L_b$ plane.

To answer our questions on interscale energy transfers at the TNTI we
need a scale-by-scale energy budget that is local in space and
time. This budget is the fully generalised K\'arm\'an-Howarth
equation, i.e. the version of the K\'arm\'an-Howarth-Monin-Hill (KHMH)
equation directly derived from the incompressible Navier-Stokes
equations for the instantaneous velocity field \cite{Duchon,Hill}
without Reynolds or other decomposition, without averaging operations,
and without assumptions about the turbulence. This is the
evolution equation for $|\delta \bm{u}|^2$, where $\delta \bm{u}
\equiv \bm{u} - \bm{u}^{\prime}$ is the difference of fluid velocities
between points $\bm{x}$ and $\bm{x}^{\prime}$, $\bm{u} \equiv
\bm{u}(\bm{x},t)$ and $\bm{u}^{\prime} \equiv
\bm{u}(\bm{x}^{\prime},t)$. It is expressed in terms of centroid
$\bm{X} = ( \bm{x} + \bm{x}^{\prime} ) / 2$, separation vector $\bm{r}
= \bm{x} - \bm{x}^{\prime}$ and time $t$ as follows:
\begin{eqnarray}
\label{eq:KHMH}
\frac{\partial}{\partial t}|\delta \bm{u}|^2 &+&
\frac{\partial}{\partial r_k} (\delta u_k |\delta \bm{u}|^2 )= -
\frac{\partial}{\partial X_k} \frac{( u_k + u_k^{\prime} ) |\delta
  \bm{u}|^2}{2} \notag \\ &-& \frac{2}{\rho} \frac{\partial}{\partial
  X_k} (\delta u_k \delta p ) + 2 \nu \frac{\partial^2}{\partial
  r_k^2} |\delta \bm{u}|^2 + \frac{\nu}{2} \frac{\partial^2}{\partial
  X_k^2} |\delta \bm{u}|^2 \notag \\ &-&
\bigg[2\nu\bigg(\frac{\partial u_j}{\partial x_k}\bigg)^2 +
  2\nu\bigg(\frac{\partial u^{\prime}_j}{\partial
    x^{\prime}_k}\bigg)^2\bigg] 
\end{eqnarray}
where $\rho$ is fluid density and $\delta p = p ({\bm x})- p ({\bm
  x}^{\prime})$ is the pressure difference across the two points.

We define an average energy over scales smaller than $r =
\vert\bm{r}\vert$ as $E({\bm X}, r,t) = {3\over \pi r^{3}}\int_{V({\bm
    X}, r)} d^{3}{\bm r} |\delta \bm{u}|^2$ where the integral is over
the volume $V({\bm X}, r)$ of a sphere of diameter $r$ centred at
${\bm X}$. $E({\bm X}, r,t)$ typically varies as $r^{2}$ for $r\le
O(\eta_{K})$ and oscillates around a constant when $r$ is so large
that fluid velocities at ${\bm x}$ and ${\bm x}^{\prime}$ fluctuate
independently. The evolution equation for $E({\bm X}, r,t)$ is
obtained by applying operation ${3\over \pi r^{3}}\int_{V({\bm X}, r)}
d^{3}{\bm r}$ to all terms in equation (1), and the left side of this
evolution equation is $\frac{\partial}{\partial t}E + {3\over \pi
  r^{3}}\int_{V({\bm X}, r)} d^{3}{\bm r}\frac{\partial}{\partial r_k}
(\delta u_k |\delta \bm{u}|^2 )$. Using the divergence theorem, this
left side's second term is proportional to a scale-space flux because
$\int_{V({\bm X}, r)} d^{3}{\bm r}\frac{\partial}{\partial r_k}
(\delta u_k |\delta \bm{u}|^2 ) = \int_{\partial V({\bm X}, r)}
d^{2}{\bm r} \delta \bm{u}\cdot\hat{\bm r} |\delta \bm{u}|^2 $, where
$\int_{\partial V({\bm X}, r)} d^{2}{\bm r}$ is an integral over the
surface of the sphere of diameter $r$ centred at ${\bm X}$ and
$\hat{\bm r} \equiv {\bm r}/r$. Simplifying further, the left side
becomes $\frac{\partial}{\partial t}E + {3\over \pi}\int d\Omega
{\delta \bm{u}\cdot\hat{\bm r}\over r} |\delta \bm{u}|^2$ where
$\Omega$ is the solid angle.
The interscale transfer rate ${3\over \pi}\int d\Omega {\delta
  \bm{u}\cdot\hat{\bm r}\over r} |\delta \bm{u}|^2$ (space-scale flux
if multiplied by $r^{3}$) vanishes at $r=0$ and tends to $0$ as $r$
grows indefinitely. At a given finite scale $r$, a scale-space flux
from large to small or from small to large scales corresponds to a
negative or positive ${3\over \pi}\int d\Omega {\delta
  \bm{u}\cdot\hat{\bm r}\over r} |\delta \bm{u}|^2$ and contributes a
growth or decrease of $E({\bm X}, r,t)$ in time. In highly
inhomogeneous/anisotropic flows, particularly in the vicinity of the
TNTI, the interscale transfer rate does not necessarily dominate the
behaviour of $E({\bm X}, r,t)$ as the terms on the right side of the
equation involve a pressure-velocity term $-\frac{\partial}{\partial
  X_k} {3\over \pi r^{3}}\int_{V({\bm X}, r)} d^{3}{\bm r}
\frac{2}{\rho} (\delta u_k \delta p )$ and a term
$-\frac{\partial}{\partial X_k} {3\over \pi r^{3}}\int_{V({\bm X}, r)}
d^{3}{\bm r} \frac{( u_k + u_k^{\prime} ) |\delta \bm{u}|^2}{2}$,
which may dominate. This latter term includes mean advection and
production terms as well as the spatial turbulent transfer rate
$-\frac{\partial}{\partial X_k} {3\over \pi r^{3}}\int_{V({\bm X}, r)}
d^{3}{\bm r} \frac{( u_k - U_{k} + u_k^{\prime}-U_{k}^{\prime})
  |\delta \bm{u}|^2}{2}$ where $U_{k}$ and $U_{k}^{\prime}$ are mean
flow velocity components obtained by averaging over time at $\bm{x}$
and $\bm{x}^{\prime}$ respectively.

A forward space-scale flux from large to small scales corresponds to
predominance of compression, $\delta \bm{u}\cdot\hat{\bm r} <0$, so
that $\int d\Omega {\delta \bm{u}\cdot\hat{\bm r}\over r} |\delta
\bm{u}|^2$ is negative. Conversely, space-scale flux from small to
large scales, corresponds to predominance of stretching, $\delta
\bm{u}\cdot\hat{\bm r} >0$, so that $\int d\Omega {\delta
  \bm{u}\cdot\hat{\bm r}\over r} |\delta \bm{u}|^2$ is
positive. Incompressibility implies $\int d\Omega \delta
\bm{u}\cdot\hat{\bm r} =0$. As fluid elements approach the TNTI
without vorticity and obtain vorticity by crossing it, their velocity
normal to the TNTI changes depending on reorientation of their motion
(which suddenly becomes vortical) and on a sudden decrease in
pressure~\cite{Reynolds,Westerweel}. If this \textcolor{black}{normal
  velocity change} helps sustain the TNTI, one may expect $\delta
\bm{u}\cdot\hat{\bm r}$ to be negative for $\hat{\bm r}$ around the
normal to the TNTI, and by virtue of $\int d\Omega \delta
\bm{u}\cdot\hat{\bm r} =0$, one may also expect
$\delta\bm{u}\cdot\hat{\bm r}$ to be positive for $\hat{\bm r}$ close
to tangent to the TNTI. Indeed, compressive relative motions in the
direction normal to the TNTI and stretching relative motions in
directions tangent to the TNTI sustain the TNTI. The question arises
whether the interscale transfers at the TNTI reflect a TNTI
self-sustaining mechanism so that $\delta \bm{u}\cdot\hat{\bm r}
|\delta \bm{u}|^2 <0$ for $\hat{\bm r}$ in the vicinity of the TNTI's
normal and $\delta \bm{u}\cdot\hat{\bm r} |\delta \bm{u}|^2 >0$ for
$\hat{\bm r}$ in the vicinity of the TNTI's tangent plane. The
question which follows is to know the sign of the resulting aggregate
interscale transfer rate $\int d\Omega {\delta \bm{u}\cdot\hat{\bm
    r}\over r} |\delta \bm{u}|^2$ at the TNTI.

To answer these questions, we start with identifying those numerical
mesh points in the $x=60L_b$ plane where $\omega$ is closest to
$\omega_{th}$. These points are on the TNTI. (Details of the procedure
to locate the TNTI are in \cite{Zhou}.) We compute ${\bm n} = {\bm
  \nabla} \omega^{2}/\vert{\bm \nabla} \omega^{2}\vert$, the unit
vector normal to the TNTI in 3D space, and select those particular
points ${\bm X}_{I}$ on the TNTI where $\omega ({\bm X}_{I} +
cL_{b}{\bm n}) > \omega_{th}$ and $\omega ({\bm X}_{I} - cL_{b}{\bm
  n}) < \omega_{th}$ for $0<c<1$. These are the points on the
interface which do not face folds over scales smaller than about 2/3
of the wake width $\delta (x)$ at $x=60L_b$ (given that $\delta
\approx 1.55 L_b$ at $x=60L_b$). We therefore limit our study to those
points on the TNTI where folds do not contribute to the TNTI's
interscale transfer properties. \textcolor{black}{With our very strict
  criterion $0<c<1$, we consider about half the points on the TNTI,
  but this proportion is much larger for a smaller range of positive
  values of $c$, in which case our analysis can be expected to carry
  over up to smaller maximum values of $r$.}

We then compute the angle $\theta_{r}$ between ${\bm n}$ and ${\bm r}$
($\cos \theta_r = {\bm n}\cdot\hat{\bm r}$) and an angle $\phi_{r}$
which locates the projection of ${\bm r}$ on the local tangent plane
normal to ${\bm n}$. Finally we calculate the conditional averages
$\langle \delta \bm{u}\cdot\hat{\bm r} |\delta \bm{u}|^2 \vert {\bm
  X}_{I}\rangle$ and $\langle \delta \bm{u}\cdot\hat{\bm r} \vert {\bm
  X}_{I}\rangle$ by averaging over 375 randomly selected instantaneous
velocity fields from 15000 time steps and over $\phi_{r}$. These
averages are conditioned on the TNTI points ${\bm X}_{I}$ and are
plotted in figure 1a,b as functions of $r_{n}/\delta$ and
$r_{t}/\delta$ where $r_n = {\bm n}\cdot{\bm r}$ and
$r_{t}^{2}=r^{2}-r_{n}^{2}$ ($\tan \theta_r = r_{t}/r_{n}$). The
results in all our figures (except figure 2b), stay very similar if
the fluid velocity ${\bm u}$ is replaced by the fluctuating velocity
${\bm u} - {\bm U}$ where ${\bm U}$ is the mean flow velocity obtained
by averaging over time.

\begin{figure}
	\includegraphics[scale=0.42,trim = 00mm 00mm 0mm 0mm, clip=true]{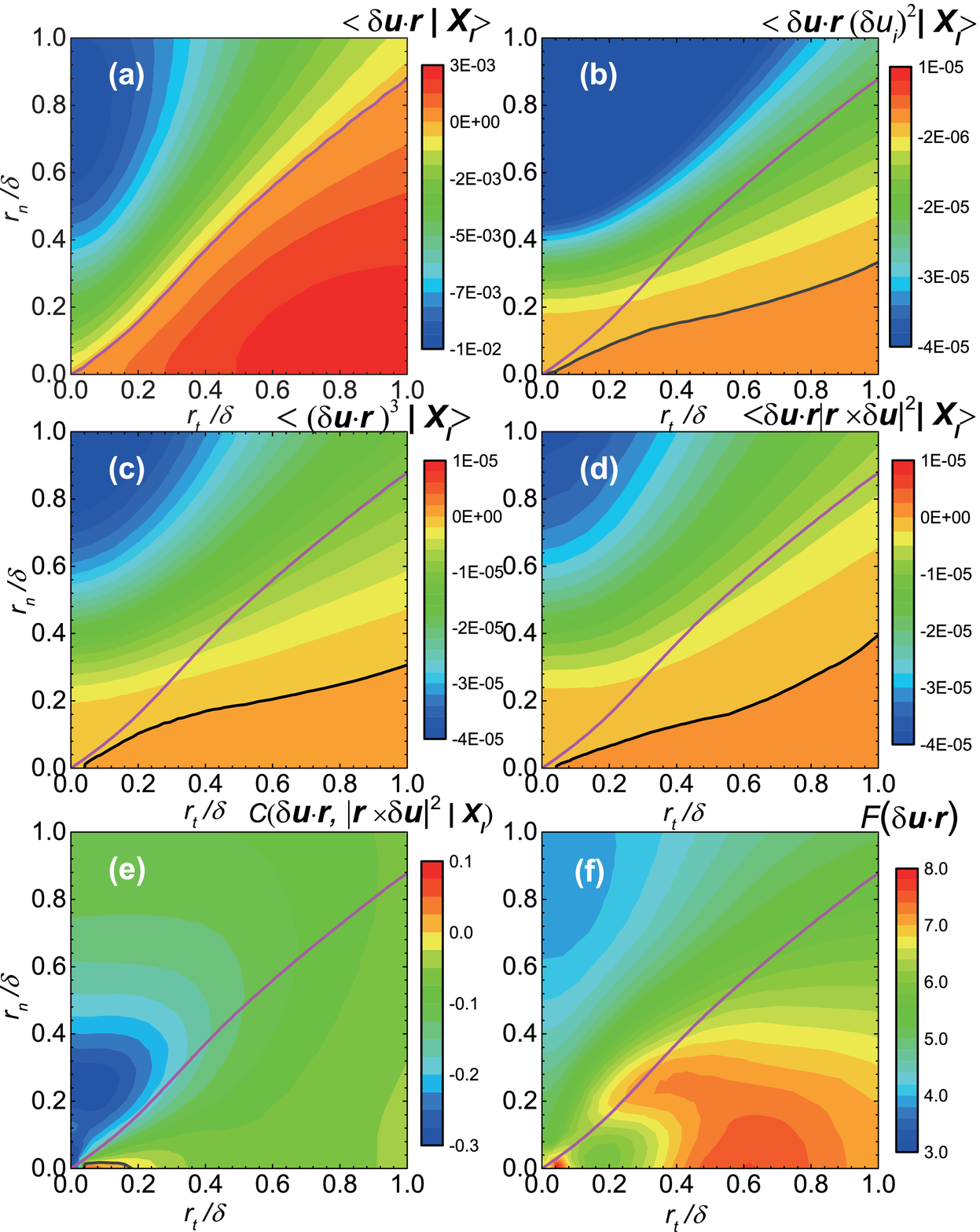}
	\caption{\label{fig1:epsart}Plots in the ($r_{n}/\delta$,
          $r_{t}/\delta$) plane (wake width $\delta$ at $x/L_{b} =
          60$) of (a) $\langle \delta \bm{u}\cdot\hat{\bm r} \vert
          {\bm X}_{I}\rangle$, (b) $\langle \delta \bm{u}\cdot\hat{\bm
            r}|\delta \bm{u}|^2 \vert {\bm X}_{I}\rangle$, (c)
          $\langle (\delta \bm{u}\cdot\hat{\bm r})^{3}\vert {\bm
            X}_{I}\rangle$, (d) $\langle \delta \bm{u}\cdot\hat{\bm r}
          |\hat{\bm r}\times \delta {\bm u}|^2 \vert {\bm
            X}_{I}\rangle$, (e) the Pearson correlation coefficient
          $C$ between $\delta \bm{u}\cdot\hat{\bm r}$ and $|\hat{\bm
            r}\times \delta {\bm u}|^2$ conditional on ${\bm X}_{I}$
          \textcolor{black}{and (f) the flatness of $\delta
            \bm{u}\cdot\hat{\bm r}$ conditional on ${\bm X}_{I}$}. The
          magenta line in plots (a) to (f) is where $\langle \delta
          \bm{u}\cdot\hat{\bm r} \vert {\bm X}_{I}\rangle =0$. The
          black line in plots (b) to (e) is where the quantity plotted
          is $0$.}
\end{figure}

Figure 1a,b shows that $\langle \delta \bm{u}\cdot\hat{\bm r} |\delta
\bm{u}|^2 \vert {\bm X}_{I}\rangle$ and $\langle \delta
\bm{u}\cdot\hat{\bm r} \vert {\bm X}_{I}\rangle$ are both negative
above the magenta line, i.e. for angles $\theta_{r}$ below about
$50^o$, and both positive below the black line in figure 1b, i.e. for
angles $\theta_{r}$ above about $70^o$-$75^o$. In this average sense,
the interscale transfers at the selected TNTI points seem to be a
reflection of compressive motions for ${\bm r}$ more or less aligned
with the normal to the TNTI and a reflection of stretching motions in
directions more or less aligned with the TNTI's tangent plane at
points ${\bm X}_{I}$. Because of these compressive and stretching
motions which sustain the TNTI, interscale transfers in the range
$r_{n}/\delta <1.0$ and $r_{t}/\delta <1.0$ are from large to small
scales for angles $\theta_{r}$ below about $50^o$ but from small to
large scales for angles $\theta_{r}$ above about $70^o$-$75^o$.

Surprisingly,
in the range of angles $\theta_{r}$ between about $50^o$ and about
$70^o$-$75^o$ where the motions are stretching on average,
i.e. $\langle \delta \bm{u}\cdot\hat{\bm r} \vert {\bm X}_{I}\rangle$
is positive, the interscale transfer is on average ``compressive'',
i.e. $\langle \delta \bm{u}\cdot\hat{\bm r} |\delta \bm{u}|^2 \vert
{\bm X}_{I}\rangle$ is negative. To better understand this
intermediate range of angles where interscale transfer is from large
to small scales even though fluid element pairs tend to separate on
average, we use the decomposition
\begin{equation}
\langle \delta \bm{u}\cdot\hat{\bm r} |\delta \bm{u}|^2 \vert {\bm X}_{I}\rangle =
\langle (\delta \bm{u}\cdot\hat{\bm r})^{3}\vert {\bm X}_{I}\rangle + \langle \delta
\bm{u}\cdot\hat{\bm r} |\hat{\bm r}\times \delta {\bm u}|^2 \vert {\bm
  X}_{I}\rangle
\end{equation}
which shows that $\langle \delta \bm{u}\cdot\hat{\bm r} |\delta
\bm{u}|^2 \vert {\bm X}_{I}\rangle$ can indeed be negative when
$\langle \delta \bm{u}\cdot\hat{\bm r} \vert {\bm X}_{I}\rangle$ is
positive if $\langle (\delta \bm{u}\cdot\hat{\bm r})^{3}\vert {\bm
  X}_{I}\rangle$ is negative enough or if $\langle \delta
\bm{u}\cdot\hat{\bm r} |\hat{\bm r}\times \delta {\bm u}|^2 \vert {\bm
  X}_{I}\rangle$ is negative enough or if both are
negative. \textcolor{black}{This decomposition states that the
  interscale energy transfer consists of a transfer of longitudinal
  energy and a transfer of rotational energy.}

Figure 1c shows that $\langle (\delta \bm{u}\cdot\hat{\bm r})^{3}\vert
{\bm X}_{I}\rangle$ has the same sign as $\langle \delta
\bm{u}\cdot\hat{\bm r} |\delta \bm{u}|^2 \vert {\bm X}_{I}\rangle$
effectively everywhere in the $(r_{n}, r_{t})$ plane.  However, figure
1d shows the same for $\langle \delta \bm{u}\cdot\hat{\bm r} |\hat{\bm
  r}\times \delta {\bm u}|^2 \vert {\bm X}_{I}\rangle$. Furthermore,
these three different conditional statistics have comparable
magnitudes effectively everywhere in the $(r_{n}, r_{t})$ plane. To
explain the sign and magnitude of the local interscale transfer rate
$\langle {\delta \bm{u}\cdot\hat{\bm r}\over r} |\delta \bm{u}|^2
\vert {\bm X}_{I}\rangle$ one therefore needs to take both $\langle
(\delta \bm{u}\cdot\hat{\bm r})^{3}\vert {\bm X}_{I}\rangle$ and
$\langle \delta \bm{u}\cdot\hat{\bm r} |\hat{\bm r}\times \delta {\bm
  u}|^2 \vert {\bm X}_{I}\rangle$ into account. For this, we use a
second decomposition, namely
\begin{eqnarray}
\langle \delta \bm{u}\cdot\hat{\bm r} |\hat{\bm r}\times \delta {\bm u}|^2
\vert {\bm X}_{I}\rangle = \notag \\ \langle \delta \bm{u}\cdot\hat{\bm r}\vert
      {\bm X}_{I}\rangle \langle|\hat{\bm r}\times \delta {\bm u}|^2 \vert {\bm
        X}_{I}\rangle +C \sigma_{CS} \sigma_{L}
\end{eqnarray}
where $\sigma_{CS}$ and $\sigma_{L}$ are standard deviations
(conditional on ${\bm X}_{I}$) of $\delta \bm{u}\cdot\hat{\bm r}$ and
$|\hat{\bm r}\times \delta {\bm u}|^2$ respectively, and $C$ is the
Pearson correlation coefficient conditional on ${\bm X}_{I}$ between
compression/stretching relative velocity $\delta \bm{u}\cdot\hat{\bm
  r}$ and $\vert {\bm L}\vert^{2}$, ${\bm L} \equiv {1\over 2}{\bm
  r}\times \delta {\bm u}$ being the angular momentum per unit mass of
the fluid elements at $\bm{x}$ and $\bm{x}^{\prime}$ with respect to
the centroid ${\bm X}$. The plot of $C$ (figure 1e) shows a small but
significant negative correlation between $\delta \bm{u}\cdot\hat{\bm
  r}$ and $\vert {\bm L}\vert^{2}$ nearly everywhere in the $(r_n,
r_t)$ plane (except in a small region along the $r_{t}/\delta$
axis). Compression at the TNTI has some positive correlation with the
square of the angular momentum relative to the TNTI, particularly for
orientations of ${\bm r}$ normal to the TNTI and up to about $45^{o}$
to that normal, and particularly for $r/\delta$ smaller than about
$0.4$.

We can now explain the sign of the local interscale transfer at points
${\bm X}_{I}$ on the TNTI. From equations (2)-(3), this sign is
determined by the signs of $\langle (\delta \bm{u}\cdot\hat{\bm
  r})^{3}\vert {\bm X}_{I}\rangle$, $\langle \delta
\bm{u}\cdot\hat{\bm r} \vert {\bm X}_{I}\rangle$ and $C$. In the
$(r_{n}, r_{t})$ plane region where $\theta_{r}$ is below about
$50^o$, all these signs are negative (figure 1a,c,e). Consequently,
the interscale transfer at our selected TNTI points ${\bm X}_{I}$ is
from large to small scales for angles up to about $50^o$ to the TNTI's
normal (figure 1b) because of the predominantly compressive motions
and because of the significant correlation of these compressive
motions with angular momentum at these angles. The compressive motions
contribute via negative values of both $\langle (\delta
\bm{u}\cdot\hat{\bm r})^{3}\vert {\bm X}_{I}\rangle$ and $\langle
\delta \bm{u}\cdot\hat{\bm r} \vert {\bm X}_{I}\rangle$ at these
orientations, meaning that extreme compressive motions are
significantly more likely than extreme stretching motions and that
motions are also compressive on average.  \textcolor{black}{The
  presence of extreme events is confirmed by figure 1f which shows
  that the flatness of $\delta \bm{u}\cdot\hat{\bm {r}}$ conditional
  on TNTI points ${\bm X}_{I}$ is larger than about $4$ throughout the
  $(r_{n}, r_{t})$ plane and increases up to values close to $8$ with
  increasing $\theta_{r}$.}
In the intermediate range of angles $\theta_{r}$ from about $50^o$ to
about $70^o$-$75^o$, the motions become stretching on average,
i.e. $\langle \delta \bm{u}\cdot\hat{\bm r} \vert {\bm X}_{I}\rangle$
becomes positive (figure 1a), but remain negatively skewed,
i.e. $\langle (\delta \bm{u}\cdot\hat{\bm r})^{3}\vert {\bm X}_{I}\rangle$
remains negative (figure 1c). In this range of angles, the interscale
transfer at the selected TNTI points ${\bm X}_{I}$ remains from large
to small scales (figure 1b) and the compressive motions remain
responsible for this forward transfer but in a different way. It is
now the fact that extreme compressive motions are significantly more
likely than extreme stretching motions and that the compressive
motions remain correlated with angular momentum which keeps this
interscale transfer flowing from large to small scales. The negative
values that these two effects contribute to $\langle \delta
\bm{u}\cdot\hat{\bm r} |\delta \bm{u}|^2 \vert {\bm X}_{I}\rangle$
(equations (2)-(3)) overcome the positive value of $\langle\delta
\bm{u}\cdot\hat{\bm r}\vert {\bm X}_{I}\rangle \langle|\hat{\bm
  r}\times \delta {\bm u}|^2 \vert {\bm X}_{I}\rangle$ in this
intermediate range of angles $\theta_{r}$ where the motions are now
stretching on average.

\begin{figure}
	\includegraphics[scale=0.50,trim = 00mm 00mm 0mm 0mm, clip=true]{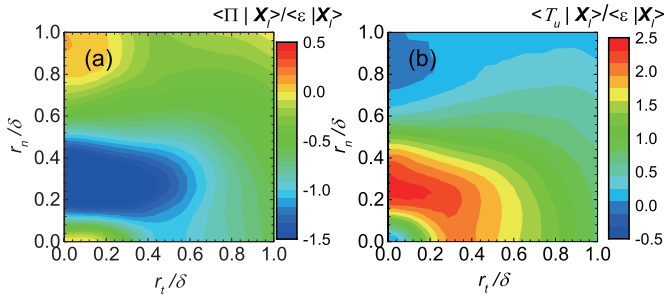}
	\caption{\label{fig2:epsart} Plots in the $(r_{n}/\delta ,
          r_{t}/\delta)$ plane (wake width $\delta$ at $x/L_{b} = 60$)
          of (a) $\langle\Pi \vert {\bm
            X}_{I}\rangle/\langle\varepsilon_{r}\vert {\bm
            X}_{I}\rangle$ and (b) $\langle T_{u} \vert {\bm
            X}_{I}\rangle/\langle\varepsilon_{r}\vert {\bm
            X}_{I}\rangle$.}
\end{figure}

As the angle $\theta_{r}$ grows beyond $70^o$-$75^o$, the motions
become predominantly stretching both on average and also in the sense
that extreme stretching events become more likely than extreme
compressive events (figure 1a,c). Furthermore the correlation between
compressive motions and angular momentum is weaker in the vicinity of
the TNTI's tangent plane. In fact, $C$ does not only take weaker
negative values but even small positive ones for some values of $r_n$
and $r_t$ (figure 1e). The result (figure 1b) is a small positive
$\langle \delta \bm{u}\cdot\hat{\bm r} |\delta \bm{u}|^2 \vert {\bm
  X}_{I}\rangle$ (equations (2)-(3)) and therefore a weak average
interscale backscatter at these orientations.

The next question concerns the sign of the time-average conditional on
${\bm X}_{I}$ of the resulting aggregate interscale transfer rate
${3\over \pi}\int d\Omega {\delta \bm{u}\cdot\hat{\bm r}\over r}
|\delta \bm{u}|^2$ at the TNTI. This conditional time-average is $12
\int_{0}^{\pi/2} d\theta_{r} \sin \theta_{r} \langle{\delta
  \bm{u}\cdot\hat{\bm r}\over r} |\delta \bm{u}|^2\vert {\bm
  X}_{I}\rangle$ and it equals ${12\over r^{3}} \int_{0}^{r} dr r^{2}
\int_{0}^{\pi/2} d\theta_{r} \sin \theta_{r} \langle\Pi \vert {\bm
  X}_{I}\rangle$ where $\Pi \equiv \Pi ({\bm X}, {\bm r}) \equiv
\frac{\partial}{\partial r_k} (\delta u_k |\delta \bm{u}|^2 )$. In
figure 2a, the plot of $\langle\Pi \vert {\bm
  X}_{I}\rangle/\langle\varepsilon_{r}\vert {\bm X}_{I}\rangle$
in the $(r_n, r_t)$ plane ($2\varepsilon_{r}$ is the sum of the
kinetic energy dissipation rates at $\bm{x}$ and $\bm{x}^{\prime}$)
shows that $\langle\Pi \vert {\bm X}_{I}\rangle$ is negative
everywhere within $0\le r_{n}/\delta \le 1.0$ and $0\le r_{t}/\delta
\le 1.0$ except in a small region at the top left of this region of
the $(r_n, r_t)$ plane where it takes small positive values. Clearly,
$\int_{0}^{\pi/2} d\theta_{r} \sin \theta_{r} \langle\Pi \vert {\bm
  X}_{I}\rangle$ is negative for all $r/\delta \le 1.0$:
inspite of the combined forward and backscatter interscale transfers
at different orientations with respect of the TNTI, the resulting
aggregate interscale transfer at the TNTI is from large to small
scales for all scales $r\le \delta$.

Spatial energy transfers turn out to be as, if not even more,
important than the interscale transfers at the TNTI, and to even
correlate with them in the $(r_n, r_t)$ plane. Compare the plot of
$\langle\Pi \vert {\bm X}_{I}\rangle/\langle\varepsilon_{r}\vert {\bm
  X}_{I}\rangle$ in figure 2a with the plot of $\langle T_{u} \vert
{\bm X}_{I}\rangle/\langle\varepsilon_{r}\vert {\bm X}_{I}\rangle$ in
figure 2b, where $T_u \equiv -\frac{\partial}{\partial X_k} \frac{(
  u_k - U_{k} + u_k^{\prime}-U_{k}^{\prime}) |\delta \bm{u}|^2}{2}$ is
the spatial non-linear energy transport rate in equation (1). The
scales and orientations where the local interscale transfer rate takes
its highest positive values correspond to those where the local
spatial transfer rate takes its highest negative values. These scales
($r_n$ in particular) are comparable to the centreline Taylor length
scale $\lambda$ ($\approx 0.24 \delta$ at $x=60L_b$). Furthermore,
the small upper left corner in the $(r_n, r_t)$ plane where $\langle
\Pi \vert {\bm X}_{I}\rangle$
is positive (and small) is also the only region in this plane where
$\langle T_{u} \vert {\bm X}_{I}\rangle$
is negative (and small). The magniture of $\langle T_{u} \vert {\bm
  X}_{I}\rangle$
is typically about twice the magnitude of $\langle \Pi \vert {\bm
  X}_{I}\rangle$.
Both interscale and spatial energy transfers result from the
Navier-Stokes convective non-linearity and this must be the root cause
of their correlation. However future investigations at much higher
Reynolds numbers should reveal the extent to which this correlation
may be due to the relatively small separation of scales between
$\lambda$ and $\delta$ in our DNS data (where the Taylor length-based
Reynolds number $Re_{\lambda} = 57$ at $x/L_{b}=60$ on the centreline)
\textcolor{black}{and/or the extent to which this correlation is an
  essential part of energy transfers in locally or statistically
  inhomogeneous situations.}

Having analysed interscale transfers in the vicinity of and relative
to the TNTI we now investigate whether the position of the TNTI
relative to the centreline $y=z=0$ affects interscale transfers and
dissipation. We calculate radial TNTI positions $R_{I} (\phi,t)$ in
the $(y,z)$ plane at $x=60L_b$ by finding the TNTI's intersections
with radial straight lines in this plane which cross the centreline's
position $y=z=0$ with an azimuthal angle $\phi$. Relatively rarely,
there are more than one intersection between the straight line and the
TNTI, in which case the recorded $R_I$ value is the largest. We define
$\Pi^a ({\bm X},r,t) = \int d\Omega \Pi$ and consider locations ${\bm
  X}$ in the $x=60L_b$ plane. For such locations, $\Pi^a ({\bm X},r,t)
= \Pi^a (R,\phi,r,t)$. We focus on $r\approx \lambda$ as the
interscale transfer rate is highest near this length-scale at the TNTI
and calculate averages over time and $\phi$ of $\Pi^a
(R,\phi,0.4L_{b},t)$ ($\lambda \approx 0.38 L_{b}$ at $x=60L_b$), the
kinetic energy dissipation rate $\varepsilon ({\bm X},t) = \varepsilon
(R,\phi,t)$ and $\omega ({\bm X},t) = \omega (R,\phi,t)$ conditional
on $R_I$.
These conditional averages (plotted in figure 3) are functions of $R$
and $R_I$. They all take their largest magnitudes when $R_I$ is
between $2\delta$ and $3\delta$, the furthest distances from the
centreline where the TNTI is found.
The actual value of $R$ does not seem to matter other than it should
not be smaller than about $\delta/5$ for these three conditional
averages to achieve high magnitudes when $R_I$ is so large. The high
magnitudes of $\Pi^a$ are negative, reflecting interscale transfers
from large to small scales on average.

\begin{figure}
	\includegraphics[scale=0.50,trim = 00mm 00mm 0mm 0mm, clip=true]{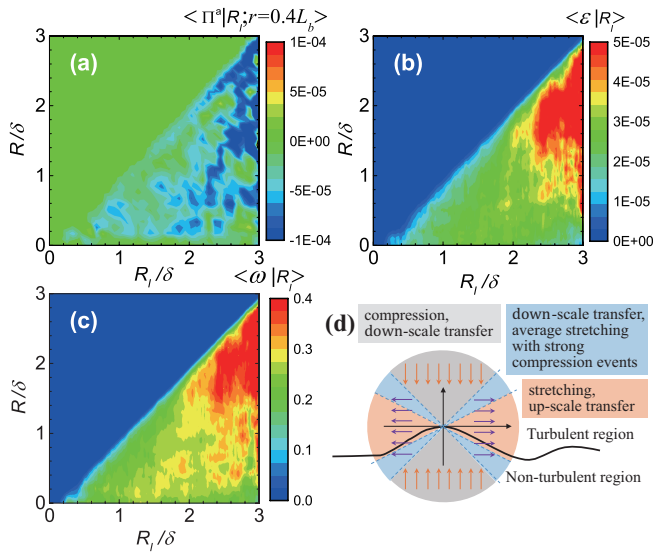}
	\caption{\label{fig3:epsart} Averages conditional on $R_I$ of
          (a) $\Pi^a$, (b) $\varepsilon$, (c) $\omega$ as functions of
          $R_I$ and $R$ and (d) \textcolor{black}{Summary schematic of
            the three intercale transfer regions at the TNTI}.}
\end{figure}

{\it Conclusion.} The most intense average interscale transfer and the
most intense average dissipation and $\omega$ occur when the TNTI is
furthest from the centreline and do so more or less uniformly all the
way from the TNTI to a finite distance from the
centreline. Occurrences of large patches of high enstrophy may
simultaneously cause the TNTI to be pushed far from the centreline and
interscale transfers to be intense and forward with high
dissipation. At the TNTI, interscale transfers are weak and backward
in directions close to the TNTI's tangent plane because of straining
motions but forward in the other directions because of compressive
motions. The interscale transfer at the TNTI is forward where extreme
compressive motions are more likely than extreme stretching motions
even when motions are stretching on average \textcolor{black}{(see
  summary schematic in figure 3d)}. A positive correlation exists at
the TNTI between compressive motions and angular momentum
magnitude. This correlation makes a forward contribution to interscale
energy transfers at the TNTI.

\begin{acknowledgments}
We were supported by the National Natural Science Foundation of China
(Nos.~91952105 and~11802133), ERC Advanced Grant 320560 and Chair of
Excellence CoPreFlo funded by I-SITE/MEL/Region Hauts de France.
\end{acknowledgments}

\nocite{*}

\bibliography{apssamp}

\end{document}